# Monitoring morphological and chemical properties during silver solid-state dewetting

M. Berni[1], I. Carrano[2], A. Kovtun[3], A. Russo[4], A. Visani[5], C. Dionigi[6], A. Liscio[7], F. Valle[6] and A. Gambardella[8,*]

[1]Dipartimento di Ingegneria dell'Informazione, Università degli Studi di Brescia, Via Branze 38 - 25123 Brescia – Italy

[2]Dipartimento di Chimica "Giacomo Ciamician", Università degli Studi di Bologna, Via Selmi 2, 40126 Bologna, Italy

[3]Consiglio Nazionale delle Ricerche, Istituto per la Sintesi Organica e la Fotoreattività, (CNR-ISOF), Via Gobetti 101, Bologna, Italy

[4]IRCCS Istituto Ortopedico Rizzoli, Clinica Ortopedica e Traumatologica II, Via di Barbiano 10/2- 40136-Bologna, Italy

[5]IRCCS Istituto Ortopedico Rizzoli, Laboratorio di Biomeccanica e Innovazione Tecnologica, Via di Barbiano 10/2- 40136 Bologna, Italy

[6]Consiglio Nazionale delle Ricerche, Istituto per lo Studio dei Materiali Nanostrutturati (CNR-ISMN), Via Gobetti 101, Bologna, Italy

[7]Consiglio Nazionale delle Ricerche, Istituto per la Microelettronica e Microsistemi, (CNR-IMM), Via del Fosso del Cavaliere 100, 00133 Roma, Italy

[8]IRCCS Istituto Ortopedico Rizzoli, Laboratorio di Nanobiotecnologie, Via di Barbiano 10/2- 40136 Bologna, Italy

*Corresponding author's present address: IRCCS Istituto Ortopedico Rizzoli, NanoBiotechnology Laboratory (NaBi), Via di Barbiano 10/2- 40136-Bologna, Italy. Tel.: +390516366852. E-mail address: a.gambardella@biomec.ior.it.

**Abstract**

Solid-state dewetting phenomenon in silver thin films offers a straightforward method to obtain structures having controlled shape or size -this latter in principle spanning several orders of magnitudes- with potentially strong interest in many applications involving high-tech industry and biomedicine. In this work nanostructured silver is deposited by pulsed electron ablation technique and its surface modified upon thermal treatments in air at increasing temperatures. Surface chemistry and morphology are then monitored simultaneously by X-ray photoemission spectroscopy and atomic force microscopy; in particular, the power spectral density of surface heights is used to analyze the alteration of morphology induced by annealing. It is shown that this approach adds a level of information about the dewetting process since it allows to separate between long- and short-range surface behavior and to retrieve statistical quantities relevant to a description of the features in view of applications. Our results are presented in the framework of a multidisciplinary approach, advantages and limits of which are deepened and discussed.

## 1. Introduction

Solid-state dewetting of a thin film is a thermally-driven process in which surface diffusion accounts for the minimization of surface energy; upon annealing below the melting point $T_m$, and depending on film thickness, the surface morphology evolves progressively through intermediate states until the formation of separate droplets/3D islands [1-3]. When deposited onto non-smooth surfaces, partially or totally dewetted films generally have topographic features that are correlated with features of the underlying surface. Thus, the use of lithographic and other patterning to modify the substrate surface or to pattern the film before dewetting can provide routes to controlled self-organization [4-6]; particularly interesting is the realization of nanoscale particle arrays for electronic and photonic devices, and complex structures [7-10]. Hence, ideally, dewetting offers the possibility to partially or totally tune up the film morphology. Nevertheless, a partially or completely dewetted surface represents the final state of a process that has in principle altered the physical and/or chemical properties of the system; this implies that for an exhaustive description of the system many parameters should be considered that are generally difficult to separate from each other. This requires to carefully characterize the dewetting system to discriminate against various possible effects that may influence the process. In this regard, silver (Ag) is an ideal candidate for an experimental study of the dewetting kinetics since it is inert in its metallic form; thus its topological characteristics can be separated by chemical properties, providing a simplified system to study. Recently, solid-state dewetting in Ag thin films has received renewed attention in relation to its potential interest for example in plasmonics, where the morphology and characteristics of surface nanostructures are strongly related to plasmon response and thus to sensitivity of devices [11, 12]. As Ag becomes biologically active when ionized, which typically occurs upon exposure to an aqueous media [13], applications of Ag dewetting in biomedicine involve the realization of structures with antimicrobial activity where the control over morphology is important in regulating the balance between cytotoxicity and antimicrobial action [14, 15]. The phenomenology of dewetting in polycrystalline thin films has been extensively studied in terms of distinct stages progressing from a) an induction period characterized by hillock formation due to thermal expansion, during which the surface is still continuouos and b) holes formation and expansion, with final film separation into isolated islands. Among many different studies, quantitative methods based upon scanning electron microscopy have been applied for retrieving surface parameters from second-order quantities such as autocorrelation function [16]. New insights were brought by techniques of scanning probe microscopy as for example in dewetting of polycrystalline films, where the role of grains with specific crystallographic orientation in the kinetics of formation of the holes was highlighted [2, 17-18]. However, the potential of such techniques to provide topographic images with high vertical and lateral resolution has not yet been fully explored to investigate the dewetting surface. Beyond simple imaging, further advances could be provided by the exploitation of scaling methods during each step of the dewetting process, aiming to gain a comprehensive geometrical description of the surface evolution before and after thermal activation [19, 20]. Note that in this context the adjective "geometrical" not only encompasses size and spacing of the features of a given surface, but it is also closely connected with its topological complexity and with the dynamics of the dewetting process. To better understand this point, let us consider first the as-deposited surface of a thin film before dewetting. Such a surface may be randomly rough, or exhibiting a quasi-periodic behavior at certain length scales depending on many factors inherent in the physical process by which it was obtained. For example, a thin film obtained by physical- or chemical- vapour- deposition may exhibit a quasi-periodic surface since these techniques are typically unstable respect to surface roughening because of shadowing [19]. To shed light on this subject, scaling methods based upon the statistical approach of power spectral density (PSD) function can be exploited. The latter represents a fast and well consolidated method to model the morphology and microstructure of thin films surface, as well as of many different surface patterns, such as functionalized nanoparticles [21], devices [22], materials for forensic sciences [23], and many others. In the context of dewetting, the question whether the PSD approach can be pursued to investigate the evolution of the film during dewetting has an apparently affirmative answer, given that the important morphological alterations inherent in the process are expected to have robust impact on the surface texture, so that they can be in principle monitored at each significant step. In this context, a very practical advantage of the PSD formalism is that it is able of comparing information from different techniques probing different ranges of spatial lengths scales; such a circumstance is potentially interesting when applied to dewetting, because the entire process may involve modification of the system size up to several orders of magnitude. This work attempts to tackle this subject applying the PSD methods to the evolution of the surface features during thermal annealing of Ag films deposited onto smooth substrates. To do this, a batch of nanostructured polycrystalline Ag films was deposited on amorphous silica substrates by pulsed electron ablation (PEA) [24], and scanning probe techniques used together with X-ray photoemission spectroscopy (XPS) to monitor simultaneously the evolution of morphology and the exposure of the substrate during thermal treatments. PSD functions were then extracted from topographic images acquired at different stages of thermal annealing. It is shown that thermal activation introduces a long-range periodicity in the system initially constituted by the randomly generated thin film surface; such periodicity characterizes the morphology evolution throughout the process of reshaping, up to final film separation into metallic islands. On the other hand, on spatial lengths smaller than the islands size it is shown that the roughness scales similarly regardless of the annealing temperature. These findings are potentially interesting as regard to control dewetting among applications where a periodicity is imposed in the system, as in templated dewetting. The physical meaning of statistical parameters related to average mutual distance and size of the surface features retrieved from the evolution of the morphology was also discussed. To this aim, a number of general remarks on a multi-scale approach to dewetting based upon PSD formalism is also provided.

## 2. Materials and Methods

### 2.1 Sample preparation and annealing

Film thickness, annealing temperature and time were established according to the general remarks of Thompson and Simrick [1, 25]. These authors varied the annealing temperature while keeping the annealing time constant, that is the approach chosen in this work aimed to display the relevant and different stages of dewetting. In this work, film thickness was chosen aiming to have zero holes in the initial stages of dewetting; the latter condition in high purity Ag films is expected to occur above ~ 150 nm at T ~ $0.2T_m$ ~ 200°C [1, 25-26]. Thicker films at this temperature are meta-stable, so that annealing at higher temperatures is expected to generate holes and finally island separation; thus, dewetting can be regulated simply by applying different annealing temperature to different batches of as-deposited samples. Therefore, for the present study samples with thickness of (175 ± 5) nm (Root mean square roughness $R_S$ ~ 3.5 nm at 10 x 10 μm$^2$) were deposited by PEA on silica substrates (10 × 10 mm$^2$, MicroFabSolutions s.r.l., Trento, Italy) as described previously [24]. The samples were separated into four groups; three of these were annealed in air using a Nabertherm (Lilienthal, Germany) furnace at rate of 5°C/min respectively at 200°C, 300°C and 600°C (± 5 °C) for 60 min, and then cooled down at the same rate; the remaining group was analyzed with no further treatments.

**2.2 X-Ray Photoelectron analysis**

All groups were analyzed by High Resolution X-Ray Photoelectron spectrum using a Phoibos 100 hemispherical energy analyzer (Specs GmbH, Berlin, Germany), in constant analyzer energy (CAE) mode, with pass energies of 10 eV. Photon source was Mg Kα radiation ($\hbar\omega$ = 1253.6 eV; P = 125 W). Base pressure in the chamber during analyses was $5 \cdot 10^{-10}$ mbar. A foil of pristine Ag (purity 99.99%, Goodfellow) after 5 min sonication in Acetone and 5 min in Isopropanol (J.T. Baker) was used to compare the level of contaminants with the deposited films. The XPS analysis provided the following values: Ag (61 ± 1)%, C 1s (29 ± 1)%, O 1s (7.0 ± 0.5)%, S 2 p (3.0 ± 0.5)%.

**2.3 Scanning Probe analysis**

All samples were imaged by AFM in tapping mode using a stand-alone microscope (NT-MDT Co., Moscow, Russia), equipped with Si cantilevers (tip curvature radius ≈ 10 nm and resonant frequency ≈ 240 kHz); the images were collected in air at room temperature. For the AD sample an Omicron UHV Scanning Tunneling Microscope (STM) in constant current mode was also used, ($V_B$ = 0.50V, $I_T$ = 600 pA, sample grounded). All the images were acquired at 512 × 512 points and unfiltered, except for a 2$^{nd}$ order levelling. 1D-Power Spectral Density (PSD) functions were extracted from several (> 10) non-overlapped surface regions using Gwyddyon software and then averaged. To avoid systematic errors due to surface anisotropy, the acquisition was carried out at different sample rotation angles.

## 2.2 Power spectral density

For a given surface image, characterized by a series of height values z(x, y), where x and y are spatial positions divided into a discrete number of steps N and M respectively (each step is equal to the resolution of the image measurement), the power spectral density (PSD) method can be used to determine how correlations in z(x, y) change as a function of length scale. For a given image with N rows and M columns the 1D-PSD or simply PSD is evaluated by means of the Fast-Fourier Transform as:

(1) $$\text{PSD} = P(K_x) = \frac{2\pi}{NMh} \sum_{j=0}^{N} |P_j(K_x)|^2,$$

where $h$ is the sampling step, i.e. the distance between two adjacent image points and $K_x$ is the vector K component along the x-axis in the reciprocal (or frequency) space. The j-th coefficient of the sum is given by:

(2) $$P_j(K_x) = \frac{h}{2\pi} \sum_{k=0}^{N} z_{kj} \exp(-i K_x kh),$$

where $z_{kj}$ is the height of the image point having coordinates (k, j). The 3D surface topography of thin films possess only statistical self-similarity, which takes place only in a restricted range of the spatial scales [27]. In most cases, randomly-generated surfaces of thin films obtained by stochastic processes such as near-to-equilibrium depositions can be conveniently treated as self-similar or self-affine surfaces. A surface is self-similar if its horizontal and vertical directions can be rescaled to obtain a new surface that is statistically identical to the original; different scale factors for vertical and horizontal rescaling characterize a self-affine surface [20, 28].

On a self-affine surface the Log-Log PSD curves exhibit typical features consisting of a plateau at low spatial frequencies (approximatively given by $\sim (R_s)^3$, where $R_s$ is the root mean square roughness of the image), and a monotonic decrease at higher frequencies with an onset at position $k_0$ (Fig. 1a); in this case, a convenient description of the spectra is often given by Gaussian or k-correlation models [20]. If the deposition process is non stochastic or, in general, the surface is not truly random, the surface will present physical correlations at certain lateral scales. Such a mounded or quasi-periodic surface exhibits frequency peaks in its PSD at certain frequencies $k_p$ corresponding to local maxima. In this case, the $(\Delta k_p = FWHM)^{-1}$ of each frequency peak corresponds to the mound size $\xi_p$, while the mutual distance $\lambda_p$ between the mounds corresponds to long-range periodic behavior $\lambda_p \sim (k_p)^{-1}$ [20]. In Fig. 1 the schematic of PSDs related to random and quasi-periodic surfaces is reported, together with the physical meaning of $\xi_p$ and $\lambda_p$. Finally, one notes that $\xi$ and $\lambda$ must satisfy the relation $\xi \leq \lambda$, because mounds are separated at least by their size. The existence of transition points in self-affine and periodic surfaces can be, in principle, used to extract characteristic dimensions such as grains or islands size; in the case of quasi-periodic surfaces, models other than Gaussian but realistic from the physical point of view such as Lorentzian can be implemented but they cannot be interpreted in a mathematic sense [29]. Nevertheless the curve slope $\gamma$ can be equally determined in mounded surfaces because in most cases the self-affine hypothesis remains valid at length scales much smaller than $\lambda$ [20]. The physical meaning of the statistic quantities derived from the spectra and characterizing the evolution of the surface will be discussed in the following.

# 3. Results and Discussion

## 3.1 SPM and XPS analysis

Fig. 2a shows an STM image of the As-Deposited (AD) Ag films, exhibiting grain-like appearance with average grain sizes of tens of nanometers. The nanometrically smooth profile of Ag is preserved at larger length scales, as evidenced by AFM images (Fig. 2b-c). As previously suggested, Ag deposition by PEA resulted in a significant reduction of the average size of the deposited particles compared to materials with lower thermal conductivity and/or density, such as many types of dielectrics [30-34]. It should be noted that the observed reduction is relevant as it is obtained under conditions of shadowing instability, which typically affect many techniques based upon physical- and chemical- vapor deposition, and promotes PEA as an efficient tool for reproducibly obtaining structures having average particulate size ranging from hundred of nanometers as for apatites [31-33] to tens of nanometers for ceramics [34]. Successively, during the sintering step, the process of surface diffusion driven by temperature tends to transform several grains into single grains, causing disruption of the nanostructure (Fig. 3a). After 60 min at 200°C, the images show the first stages of formation of the hillocks (induction period) whereas the surface is still continuous; the alteration of the morphology is accompanied by a roughness increase from $R_s \sim 3$ nm at (10 x 10) $\mu m^2$ scan size of the AD sample to $(40 \pm 2)$ nm of the 200°C sample evaluated over the same area.

A further increase of temperature promotes the formation of micrometer-sized holes, which is preliminar to the final separation of 3D islands at temperatures above 300°C, with exposure of the silica substrate. The roughness increases to (290 ± 30) nm and (350 ± 30) nm for the 300°C and 600°C samples respectively, according to the abrupt restructuring mechanism occurred (Fig. 3b-c). The three temperatures adopted identify respectively induction, hole formation and eventually ligament breakup which corresponds to 3D islands formation of the completely dewetted state. The percentage of surface of the substrate exposed as a function of temperature is evaluated by image analysis and reported in Table I together with the results of the XPS analysis.

Core levels related to AD and 200°C samples contain mainly Ag, C, O, in the percentage indicated in the corresponding Table I. These values are similar to those obtained on a pure Ag foil (see Section 2.2); contaminants C and O are typically present in first nanometer(s) deep layers of samples exposed to atmosphere (XPS sampling depth ~ 3 nm). Note that the simultaneous increase of both Si 2s and O 1s signals detected at 300°C and 600°C can be associated to the exposure of the substrate. In Fig.4b, peaks related to Ag 3d are shown. The two peaks at 368.2 eV and 374.2 eV correspond to Ag $3d_{5/2}$ and Ag $3d_{3/2}$ respectively, in close agreement with literature. Note from Fig. 4c that at all the temperatures films exhibit metallic Ag ($Ag^0$), as confirmed by the Binding Energy (BE) and Auger Modified Parameter (AP) value of 726.1 eV [35]. An apparent shift in signal of the Ag 3d was found at 600°C (see Fig. 4b); such phenomen is common for non homogeneous metal-insulator samples, when the surface is not equipotential, as expected for isolated island on insulating silica surface [36]. The metallic nature of the sample annealed at 600°C can be determined by the AP = (725.9 ± 0.2) eV, since it is independent of BE calibration. In summary, the findings of scanning probe microscopy and XPS analysis agree well; within the instrumental limits, the thermal treatments do not significantly affect the purity of the films in air exposure in the range of temperature considered.

## 3.2 Power Spectra analysis

Let us now turn our attention to the PSDs extracted from the topographic images and corresponding to the different morphologies originated by thermal annealing. These curves were reported separately, for sakes of clarity, in the two panels of Fig. 5 a-b. The AD sample (full dotted curve in Fig. 5a) showed self-affine behavior, with an onset corresponding at $\xi_0 = (k_0)^{-1} \sim$ 100 nm between the frequency-independent behavior at low frequency and the self-similar roughness at high frequency. A slight dependence on low frequencies is due to the occasional presence of a few nm's tall droplets, as those seen in Fig. 2b. In first approximation, the value $\xi_0$ can be associated to the average grain size, which can be regarded as unique characteristic length of the system, i.e. its correlation length. It is found that this value overestimates the grain size with respect to STM; this is due to a number of reasons inherent in comparing results from different techniques, plus relevant tip convolution effects [37]. Moreover, as highlighted previously [24], PSD functions extracted from STM images exhibited periodicity peak(s) (see inset of Fig. 5a), suggesting that the much higher spatial resolution of STM compared to AFM is recommended for evaluating the local slope $\gamma_{AD}$ and preventing from too arbitrary conclusions. Therefore, it is remarked that the AD surface *appears* self-affine when imaged on length scales typical of AFM, but actually only local self-affinity is found, which takes place on scales much smaller than the grain size. The mean value of the slope extracted from STM images was $\gamma_{AD} \sim -3.5$, as reported in Table II.

The spectra related to the annealed samples are dramatically different compared to AD (Fig. 5a-b); at 200°C the onset at high frequency disappears together with the nanostructure, and a constant-slope region appears on a much larger interval of spatial scales; moreover, peaks at low-frequency appear. These peaks can be generically associated to the new characteristic lengths - generated by the thermally activated diffusion process- or to a combination of these lengths [38]; note that above 200° the spectra must be retrieved from larger (50 x 50 $\mu m^2$) images to guarantee good sampling. The constant slope suggests local self-affine behavior at spatial scales much smaller than the average size of the features originated by annealing. This behavior does not contradict a probabilistic character of the process of nucleation and growth of holes, nor any deterministic contribution given by the presence of discontinuities [26, 39]: when observed from the length scales adopted here, the submicroscopic features appear randomly distributed over the surface and, in a certain sense, no detectable changes will affect it further with temperature increase. However, from a certain moment on, it is likely that the existence of initiation sites in correspondence of grain-boundaries or inhomogeneities introduces a characteristic length in the system given by their average mutual distance between these sites; this distance is always expected to increase, because with temperature increase the number of sites selected to become separated islands becomes gradually smaller.

Let us turn to quantitative considerations: it is believed that the considerations made at the beginning of this section for the AD sample do not apply to thermally-treated samples, because these latters exhibit a local self-affinity which spans across spatial lengths well above the instrumental limits in terms of tip convolution and/or spatial resolution. This fact is obviously a consequence of the increase of the system size with temperature. The values of the negative curve slope $\gamma$ differ significantly (~ 20%) from the AD compared to the value of ~ -4.3 found in the three thermally treated samples, despite of the severe alteration of the morphology they were subjected to. Since self-similar surfaces can be characterized by their fractal dimension, and that the latter is connected to the PSD of a given surface by an inverse power law, some authors uses the 1D-PSD to extract the fractal dimension of the system directly from AFM images. In this respect, the observed change of slope from AD to thermally treated samples might be ascribed to an increase of the fractal dimension from a value ~2 related to the nanometrically smooth AD surface to a value closer to 3 which would correspond to a rougher dewetted surface. This apparently reasonable observation is, however, too simplistic because the functional form of the PSD depends upon the algorithm used for its calculation, and for 1D-PSD applied to scanning probe images a direct estimation of scalar quantities by power (fractal) law may be not rigorous [40, 41]. Thus, discussion about PSD slope is here limited to qualitative considerations, according to other authors [42, 43]. In a previous study, an important variation was found in the PSD slope before and after depositon by PEA of ceramic thin film onto a nanometrically smooth substrate. Thereafter, the slope was observed to vary only slightly during deposition, despite of the corresponding increase of roughness (up to ~ 0.1 μm) with thickness [34]. It is likely that the "stepped" change of $\gamma$ from AD to annealed samples reflects the important and irreversibile change of the topological nature of the surface due to thermal activation; once the process is activated, microscopic changes do not affect importantly the roughness, which scales the same way in the annealed samples regardless the islands size. It should be also stressed that the fitting procedure to evaluate $\gamma_{200°C}$ could be considered more affordable than with $\gamma_{300°C}$ and $\gamma_{600°C}$ because these latter were taken under strongly inhomogeneous conditions, given that contributions of the smooth silica regions and micrometers' tall dewetted Ag islands are combined together into PSDs. This fact has important consequences when attemping to extract characteristic distances of the surface. For the annealed samples, the values of the scalar quantities $k_p$ and $\Delta k_p$ were retrieved directly from the spectra, and their statistical errors propagated to reciprocals $\lambda_p$ and $\xi_p$ (Table II). The increase of $\lambda_p$ with temperature reflects well the peaks shift resulting from islands formation and separation, as encompassed by Fig. 3; as a local curve maximum, it is related to the average distance between certain spatial regions which delimitate the local maxima of z(x, y), i. e., to some extent, the centers of the islands. However, it is underlined that the important difference between topographies at 200°C and 300°C does not allow to connect clearly the respective periodicities in the context of selection of specific sites of accumulation outlined above; this would require further studies focused on intermediate states. Above 300°C, $\lambda_p$ does not increase further as expected at the end of the dewetting process, and the obtained values are supported by the AFM findings quite well (see Fig. 3). Note that $\lambda_p$ values should be consistent with the average distance between first neighboroughs F obtained via software. However, since the threshold of minimum height in the image is selected quite arbitrarily, the uncertainty in F can become unacceptable when islands are not well separated (as during the induction period) and/or not perfectly round-shaped. In other words, F approaches to $\lambda_p$ only when taller islands are considered, that is consistent with the observation that the central value of the peak $k_p$ is related to features which most importantly affect the global roughness. The condition $\Delta k_p > k_p$ is fully satisfied only at 200°C, where $\Delta k_p$ reflects the peak spread corresponding to the induction period when the surface texture is still continuous and taller features, or hillocks, whose lateral size is roughly given by the value of $\xi_p$, have not yet merged into bigger islands. When temperature increases, the contribution of the smooth silica regions (with z(x, y) ~ 0) to the Gaussian form of Eq. 1 affects the normalization of PSDs of the 300°C- and 600°C- samples. Thus, the corresponding $\xi_p$ values cannot be considered as representative of the average island size. More significant results are expected from dewetting onto corrugated surfaces or, more specifically, patterned surfaces. Often, a manyfold distribution of distances may cause an apparent superposition between lengths apparently not truly connected each other, and PSDs could not be adequate to separate properly these lengths and their spatial distribution by simple fitting procedures [30, 38]. However, if an initial periodic perturbation is imposed to control breakup and the subsequent morphological evolution, as in templated dewetting, the frequency corresponding to its characteristic length scale would be dominant in the PSD and could be used to assess in quantitative terms a required final configuration of the system, or its self-organization [39]. Thus, further contributions on this specific subject should include numerical procedures to deconvolute multi-peaked PSD in order to separate possible different contributions from different topographic features. Finally, it is intuitive that for an affordable estimation of scalar quantities from PSD, included the case in which fitting procedures are used, it is necessary that the image size be selected aiming to obtain the highest possible spatial/frequency resolution (this also contributes to overcome possible effects of tip convolution). This may be important to deconvolute multiple peaks from PSD at least in the case in which they result well separated [24]. Not secondarily, it is straightforward that the image should contain a number of relevant features (such as the islands in the case of dewetting) capable to contribute significantly to the statistics of the PSD. This last point is not obvious, since in phenomena such as dewetting the increase of height along the z-axis may approach the instrumental limits, making difficult to track different length scales to determine different scalar parameters.

**Conclusions**

The alteration of morphology during thermal annealing in air of nanostructured Ag films was studied by means of scanning probe techniques at the different scales required by the progressive evolution of the features formed during the process. At the same time, XPS was used to support quantitatively the progressive separation between the metal and the substrate. The presented approach is interesting to gain further information about the topological complexity of the dewetting process, which ultimately reflects the important changes occurring in the physical nature of the surface.

On this basis, the different stages of dewetting consisting in thermal induction, island formation and final separation of 3D islands were imaged and PSD method was exploited for investigating the effect of temperature on the surface texture.

From a physical point of view, the induction period is dramatic since the disruption of the film nanostructure driven by the temperature -i. e. the process by which grains with average sizes of tens of nanometers merge into single bigger grains- introduces a periodicity in the system resulting in a frequency peak in the PSD; at higher temperatures, this long-range behavior is likely associated to the characteristic lengths brought into system by the mutual distance between the accumulation sites selected to become isolated islands, and evidences a crossover between the initial randomly-generated surface and the quasi-periodic surface originated via annealing. Transition points in the PSDs can be used via fitting procedure to retrieve quantitative parameters during the evolution of the surface; on the basis of the procedure discussed in the present work, it is believed that the formalism used here is promising in applications where a periodicity is imposed to system, such as in templated dewetting. Nevertheless, the results obtained show that further studies are desirable to investigate via PSD method a number of intermediate states during the induction period to elucidate the process of selection of sites and to develop numerical procedures for determining the different characteristic lengths of the surface when the latter has achieved strongly periodic self-organization.

**Acknowledgements**

This work has received funding from INAIL-Istituto Nazionale Assicurazione Infortuni sul Lavoro (National Institute for Insurance against Accidents at Work) [CUP = E55F16000020005] provided by the Rizzoli Orthopedic Institute, and from the European Union's Horizon 2020 research programme GrapheneCore2 785219 – Graphene Flagship. The authors acknowledge Dr. Massimiliano Cavallini for his kind support to the scientific content of the present work, all the staff working at the facilities of ISMN-CNR@spmlab and Mr. Federico Bona for his kind technical support.

Table I- Percentage of elements resulting from the XPS carried out after the different annealing temperatures

| Sample | Substrate area exposed [%] | Ag $3d_{5/2}$ [%] | C 1s [%] | O 1s [%] | Si 2s [%] | S $2p_{3/2}$ [%] | Cl 2p [%] | I 2p [%] |
|---|---|---|---|---|---|---|---|---|
| AD | - | 53.6 ± 0.9 | 37.5 ± 0.7 | 6.1 ± 0.3 | 1.1 ± 0.1 | 1.7 ± 0.1 | - | - |
| 200°C | < 1 | 50.6 ± 0.9 | 38.0 ± 0.7 | 8.7 ± 0.4 | 1.3 ± 0.2 | - | 2.7 ± 0.2 | - |
| 300°C | 28.5 ± 4.4 | 30.3 ± 0.7 | 25.9 ± 0.5 | 26.3 ± 0.5 | 15.9 ± 0.3 | - | 1.5 ± 0.1 | 0.17 ± 0.03 |
| 600°C | 60.0 ± 3.3 | 9.4 ± 0.7 | 17.9 ± 0.5 | 42.6 ± 0.5 | 29.0 ± 0.4 | - | 1.1 ± 0.1 | - |

Table II- Statistical parameters retrieved from PSD functions related to the different annealing temperatures

| Sample | $\gamma$ | $k_p$ [μm$^{-1}$] | $\lambda_p$ [μm] | $\Delta k_p$ [μm$^{-1}$] | $\xi_p$ [μm] |
|---|---|---|---|---|---|
| AD | -3.49 ± 0.05 | - | - | - | - |
| 200°C | -4.38 ± 0.04 | 0.564 ± 0.06 | 1.77 ± 0.19 | 0.687 ± 0.149 | 1.46 ± 0.31 |
| 300°C | -4.16 ± 0.04 | 0.227 ± 0.02 | 4.40 ± 0.39 | 0.199 ± 0.04 | 5.02 ± 1.01 |
| 600°C | -4.39 ± 0.05 | 0.238 ± 0.01 | 4.20 ± 0.18 | 0.162 ± 0.01 | 6.17 ± 0.38 |

Figure 1- a) PSD related to a randomly generated and self-affine surface and b) a quasi-periodic surface. c) z-profile of a quasi-periodic (mounded) surface with evidenced the distance between the mounds $\lambda_p$ and the mound size $\xi_p$ related to the corresponding PSD peak at $k = k_p$.

Figure 2- a) (300 x 300) nm$^2$ STM topography of the nanostructured Ag surface before thermal treatments (AD sample) and (b) AFM (10 x 10) µm$^2$ image taken on the same surface. Roughness profile is indicated by dotted white line and reported in c).

Figure 3- a) From top to bottom: 50 x 50 µm$^2$ and zoomed 10 x 10 µm$^2$ images showing hole induction b) hole formation and c) final separation of 3D islands.

Figure 4- a) XPS survey spectrum taken at the different annealing temperatures. (b) Detail of the 3d 5/2 and 3d 3/2 peaks of Ag. (c) Auger MVV peaks for the estimation of the Auger modified parameter.

Figure 5- a) PSDs related to AD and 200°C samples evaluated from 10 x 10 µm² images. In the inset, PSD related to AD sample evaluated from 300 x 300 nm² STM images. b) PSDs related to 300°C and 600°C samples evaluated from 50 x 50 µm² images.

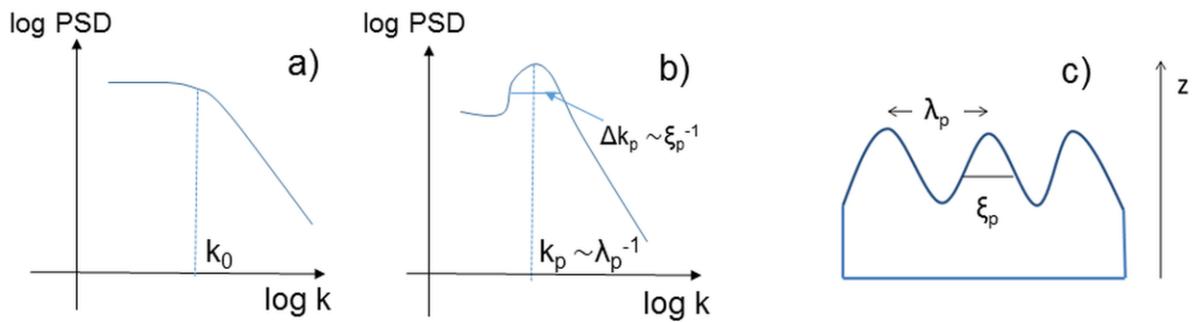

Figure 1

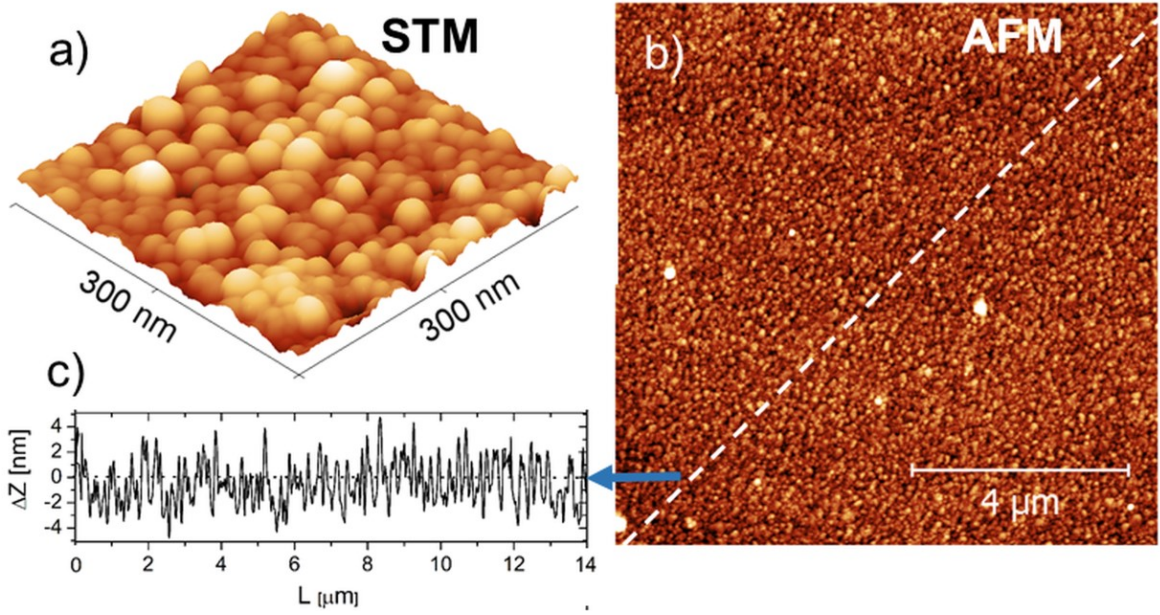

Figure 2

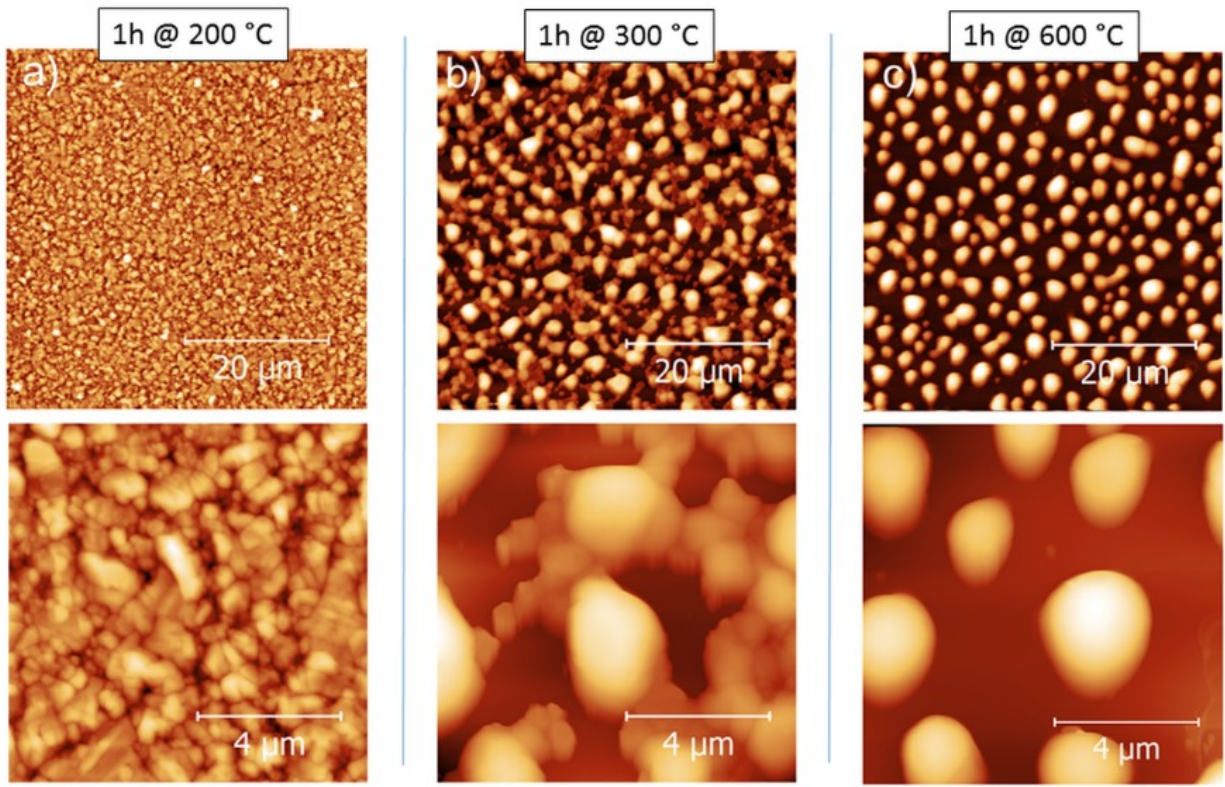

Figure 3

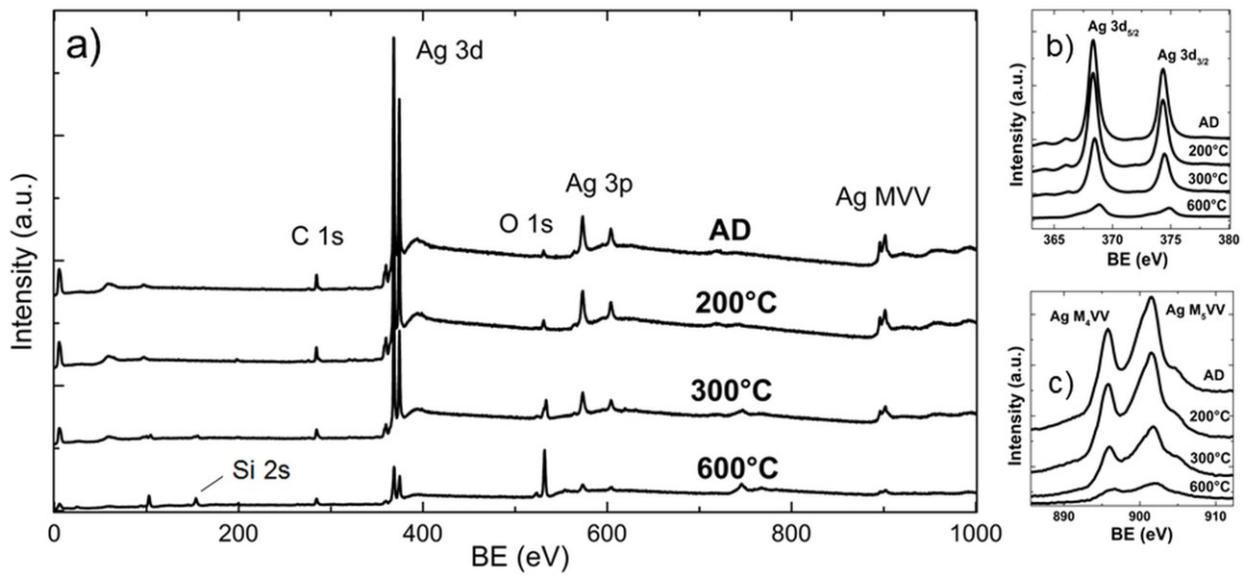

Figure 4

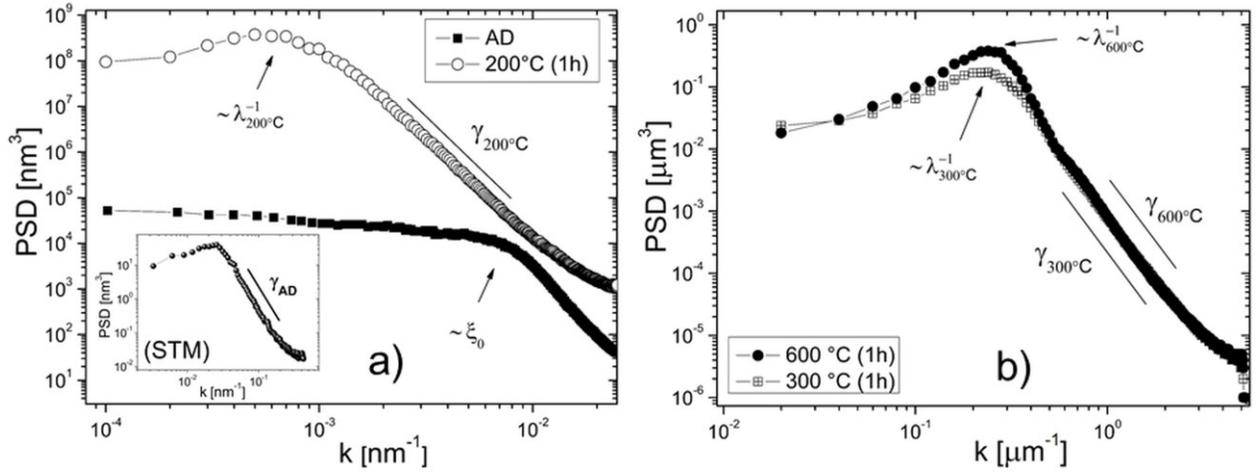

Figure 5